\documentclass[journal]{IEEEtran}
\usepackage{amsmath}
\usepackage{amssymb}
\usepackage{amsfonts}
\usepackage{amsthm}
\usepackage{graphics}
\usepackage{graphicx}
\usepackage{caption}
\usepackage{subcaption}
\usepackage{epstopdf}
\newtheorem{theorem}{Theorem}
\newtheorem{lemma}{Lemma}
\usepackage[ruled,vlined]{algorithm2e}

\begin{document}
\title{Joint Resource Allocation for OFDMA based Overlay
 Cognitive Radio Networks under Stochastic Rate Constraint}
\author{Ayush~Kumar}
        
\maketitle

\begin{abstract}
This work presents joint subcarrier, power and bit allocation schemes for multi-user OFDM based overlay cognitive radio under constraints on primary user throughput loss and total secondary user transmitted power. Closed form for expressions for the optimal power and bit allocations are initially derived. Subsequently, optimal and suboptimal algorithms to realise the joint subcarrier, power and bit allocations are introduced. Numerical simulations are used to evaluate the performance of the joint resource allocation algorithms.
\end{abstract}


\IEEEpeerreviewmaketitle

\section{Introduction}
Cognitive Radio (CR) \cite{MITOLA00} is a radio device that intelligently senses the communication environment, analyses the changes and adapts itself accordingly. There are two types of users in a cognitive radio system, primary users (PUs) and secondary users (SUs). PUs are licensed to operate in a given spectrum, while SUs are unlicensed and can share the spectrum with PUs in three different ways which form the paradigms of cognitive radio: overlay, underlay and interweave. In overlay cognitive radio, SUs transmit even when the spectrum is occupied by PUs while in underlay cognitive radio, SUs transmit only when the spectrum is unoccupied. When SUs transmit in spectrum bands interweaved with the bands occupied by PUs, it is called interweave cognitive radio.


Moreover, OFDM has been identified as a suitable modulation technique for cognitive radio systems \cite{WEISS04} since it allows  high flexibility with respect to the transmitted signal's spectral shape so that it only fills the spectral bands unoccupied by a licensed user. This ensures that the mutual interference between the licensed system and the unlicensed system is minimized. OFDM offers several other advantages such as high spectral efficiency, frequency flat fading in subcarriers, robustness against Inter Symbol Interference (ISI) and efficient implementation using Fast Fourier Transform (FFT) algorithms. However, the subcarriers in an OFDM system may experience varying channel conditions with time. Hence adaptive bit and power allocation algorithms are needed
that assign appropriate transmission power and bits to each subcarrier so that the overall ergodic capacity of the system is maximized. Also, in a multi-user communication system, there exists a multi-user diversity which arises due to independent fading channels across the multiple users. Subcarrier allocation is a technique that takes advantage of this channel diversity across users to improve overall system throughput.

We now go through the existing literature for subcarrier, power and bit allocation in OFDM based cognitive radio systems.
In \cite{CIOFFI00}, Cioffi et al. form a multi-user convex optimization problem to find the optimal subcarrier allocation in a multiuser OFDM based system and proposed a low-complexity adaptive subcarrier allocation algorithm. A joint overlay and underlay power allocation scheme is investigated by the authors in \cite{DUVAL10} by maximizing the total capacity of cognitive radio while maintaining a total power budget and keeping the interference introduced to the primary user band below a threshold. In \cite{DUALPA10}, the power allocation problem is formulated as a convex optimization problem and dual decomposition method is used to obtain the optimal power allocation. Kang et al. \cite{PRIMPROT10} propose a rate loss constraint for primary transmission protection and derive optimal power allocation under that constraint using dual decomposition. In \cite{TANG10}, integer linear programming based adaptive bit loading and subcarrier allocation techniques are proposed which consider interference leakage to and from multiple primary users and secondary users to optimise the throughput of the secondary network while keeping the interference to primary network below a threshold. The authors in \cite{DAWEI10} present a new bit allocation scheme for cognitive OFDM systems based on the margin adaptive principle in which the overall transmission power is minimized by placing constraints on the fixed data rate and bit error rate. Nadkar et al.\cite{NADKAR10} propose a bit allocation algorithm to be used in an OFDM based cognitive relay network using a two pass algorithm.

In this paper, we mathematically characterize the joint subcarrier, power and bit allocation problem in a multiuser OFDM based overlay cognitive radio system under constraints on PU throughput loss and total SU transmitted power. Using Lagrange multipliers method and dual decomposition, we attempt to solve the resulting non-convex optimization problem and derive closed form expression for the power and bit allocation. Further, we propose optimal and sub-optimal algorithms to realise the joint subcarrier, power and bit allocations. Finally, the performance of the proposed resource allocation algorithms is evaluated using MATLAB simulations.

\section{System Model}
Consider an OFDM based overlay cognitive radio system with $N$ subcarriers allocated to $M$ PUs of the primary system. The secondary system consists of $K$ SUs sharing the same $N$ subcarriers with the primary system. For the $i^{th}$ subcarrier, let the instantaneous channel gains of the primary link, the secondary link, the link between  PU transmitter and SU receiver and the link between SU transmitter and PU receiver be denoted by $H_{i}^{pp}$, $H_{i}^{ss}$, $H_{i}^{ps}$ and $H_{i}^{sp}$ respectively. It is assumed that the SUs use a fraction of its transmit power to relay PU transmissions so that the maximum loss in PU throughput (in bits/s) due to SU interference is upper bounded. Also, the interference link $H_{i}^{sp}$ is assumed to be weak, so the PUs treat it as noise.

\begin{figure}
\centering
\includegraphics[width=90mm,height=60mm]{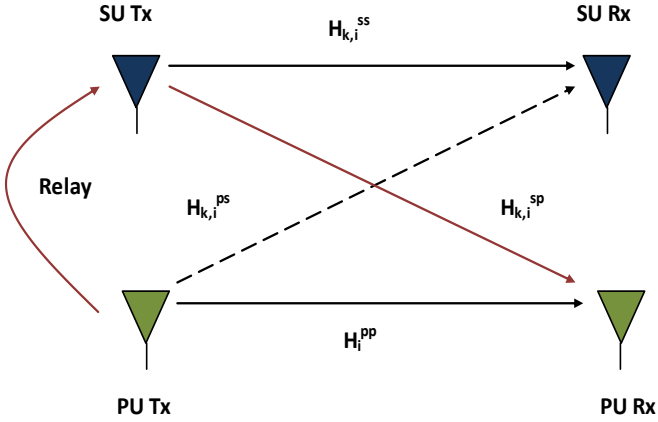}
\caption{Channel model for subcarrier $i \in U_k$} 
\end{figure}


\subsection{PU throughput}

The maximum throughput of $j^{th}$ PU in absence of the secondary system is given by
\begin{equation}
R_{j}^{p}=\sum_{i \in \Omega_j} \log_2 \left(1+\frac{|H_i^{pp}|^2 T_i}{N_o}\right)
\end{equation}
where $\Omega_j$ is the set of subcarriers assigned to $j^{th}$ PU and $T_i$ is the power allocated to the $i^{th}$ subcarrier by PU and $N_o$ is the AWGN noise power in a subcarrier.
The maximum throughput of $j^{th}$ PU with SU relaying is given by \cite{VISW09}
\begin{equation} \label{primthput}
R_{j}^{s}=\sum_{k \in \Psi} \sum_{i \in \Omega_j} \log_2 \left(1+\frac{(|H_i^{pp}| \sqrt{T_i}+|H_{k,i}^{sp}| \sqrt{\alpha_k P_{i}})^2}{N_o+|H_{k,i}^{sp}|^2 (1-\alpha_k) P_{i}}\right)
\end{equation}
where $\Psi = \{k \in (1,2,\dots,K)\, | \, \Omega_j \cap U_k \neq \emptyset\}$, $U_k$ is the set of subcarriers allocated to the $k^{th}$ SU, $P_i$ is the power allocated to $i^{th}$ subcarrier by SU, and $\alpha_k$ is the fraction of transmission power used for relaying by $k^{th}$ SU.

\subsection{PU Activity Model}
The arrival-departure process of PUs can be reliably modelled using a two-state Markov chain model \cite{HORVATH11}. The ON state indicates the presence while the OFF state indicates the absence of $j^{th}$ PU. The corresponding state transition matrix can be written as,
\begin{eqnarray*}
\mathbf{P}=
	\begin{pmatrix}
	1-\alpha_j & \alpha_j \\
	\beta_j & 1-\beta_j \\
	\end{pmatrix}.
\end{eqnarray*}
where $\alpha_j=Pr(S_{n+1}=OFF | S_n=ON)$ and $\beta_j=Pr(S_{n+1}=ON | S_n=OFF)$. Over time, the Markov chain converges to the steady state distribution expressed as $\pi=(p_{j,0} \, p_{j,1})$ which is obtained by solving the equation,
\begin{eqnarray}
\pi \mathbf{P} = \pi
\end{eqnarray}
where $p_{j,0}$ and $p_{j,1}$ represent the steady-state probability for the OFF/ON states of the $j^{th}$ PU respectively.
\begin{eqnarray}
p_{j,0}=\frac{\beta_j}{\alpha_j+\beta_j} \\
p_{j,1}=\frac{\alpha_j}{\alpha_j+\beta_j}
\end{eqnarray}

\section{Joint subcarrier, power and bit allocation} \label{glrtbaseddet}

The maximum number of bits that can be loaded into the $i^{th}$ subcarrier is given by \cite{CIOFFI91}
\begin{equation} \label{bitloading}
b_{k,i}=\log_2 \left(1+\frac{{SINR}_{k,i}}{\Gamma}\right)\,\,\forall \,i \in U_k, k \in \{1,2,\dots,K\} 
\end{equation}
where ${SINR}_{k,i}$ is the signal to interference-plus-noise ratio and $\Gamma$ is the SNR gap which is calculated by \textit{gap approximation formula} \cite{CIOFFI91}, based on target bit error rate (BER).
SNR gap ($\Gamma$) represents how far the system is from theoretical capacity of $\log_2 (1+SNR)$ for AWGN channel.

Assuming rectangular M-QAM modulation and ideal coherent phase detection, the SNR gap is given by \cite{CIOFFI91},
\begin{equation} \label{snrgap}
\Gamma \geq \frac{1}{3}\left[Q^{-1}(P_e/4)\right]^2
\end{equation}
where $Q^{-1}$ is the inverse of the Q-function defined as,
\begin{equation}
Q(x)=\frac{1}{\sqrt{2\pi}}\int_x^{\infty}e^{-t^2/2} \,dt
\end{equation}
If instead we use MPSK modulation \cite{MPSK06}, the SNR gap is given by,
\begin{equation}
\Gamma^{*} \geq \left[\frac{Q^{-1}(P_e/2)}{\pi \sqrt(2)}\right]^2
\end{equation}

We will use \eqref{snrgap} with equality sign for SNR gap approximation.
For the system model under consideration, \eqref{bitloading} can be re-written as,
\begin{equation} \label{bitloadingexp}
b_{k,i}=\log_2 \left(1+\frac{|H_{k,i}^{ss}|^2 P_{i}(1-\alpha_k)}{\Gamma(N_o +\bar{J}_{k,i})}\right)
\end{equation}
where $\bar{J}_{k,i}$ is the expected interference introduced by $j^{th}$ PU into the $i^{th}$ subcarrier given by 
\begin{eqnarray}
\bar{J}_{k,i}&=&p_{j,1} \times 0+p_{j,1}|H_{k,i}^{ps}|^2 T_i \nonumber \\
&=&p_{j,1}|H_{k,i}^{ps}|^2 T_i, \, \forall \, i \in \Omega_j
\end{eqnarray}
We consider interference contribution to $\bar{J}_{k,i}$ from PU transmission in subcarriers other than the $i^{th}$ subcarrier to be negligible.

Since in the overlay paradigm, the SU transmits in a subcarrier regardless of whether PU is already present, we propose a minimum transmission rate for the corresponding PU so that it need not worry about performance degradation. Hence, the stochastic rate constraint for the $j^{th}$ PU is given by
\begin{eqnarray}
\bar{R}_{j}^{s} \geq R_j
\end{eqnarray}
where $\bar{R}_{j}^{s}=p_{j,0} \times 0+p_{j,1}R_{j}^{s}=p_{j,1}R_{j}^{s}$ is the expected $j^{th}$ PU transmission rate and $R_j$ denotes the minimum required rate. Applying a total power constraint $P_t$, the joint bit, power and subcarrier allocation optimization problem maximizing bit rate can be written as,
\begin{equation*}
\max_{U_k,P_i} \sum_{k=1}^K \sum_{i \in U_k} b_{k,i}
\end{equation*}
subject to
\begin{equation*}
\bar{R}_{j}^{s} \geq R_j, \forall j \in \{1,2,\dots,M\}
\end{equation*}
\begin{equation*}
\sum_{i=1}^N P_{i} \leq P_t 
\end{equation*}
\begin{equation}
b_{k,i} \in Z_{+}, \forall i \in U_k, k \in \{1,2,\dots,K\}
\end{equation}
Due to the integer bit constraint, the above optimization problem turns out to be a \textit{combinatorial} one \cite{RMLOAD08}. To make it mathematically tractable, we relax the integer bit constraint to
\begin{equation}
b_{k,i} \geq 0 
\end{equation}
Using \eqref{bitloadingexp}, the throughput maximization problem can be restated as 
\begin{equation*}
\max_{U_k,P_i} \sum_{k=1}^K \sum_{i \in U_k}\log_2 (1+ s_{k,i}P_{i})
\end{equation*}
subject to
\begin{equation*}
\bar{R}_{j}^{s} \geq R_j, \forall j \in \{1,2,\dots,M\}
\end{equation*}
\begin{equation*}
\sum_{i=1}^N P_{i} \leq P_t 
\end{equation*}
\begin{equation} \label{jointoptprob}
P_{i}\geq 0, \forall i \in \{1,2,\dots,N\} 
\end{equation}
where $s_{k,i} = {|H_{k,i}^{ss}|^2(1-\alpha_k)}/{\Gamma (N_o+J_{k,i})}$.
To perform the subcarrier allocation feasibly, we assume equal power allocation ($P_{eq}$) in all subcarriers though the algorithm would be suboptimal. However, we still have to satisfy the total power constraint, so $P_{eq}$ is given by,
\begin{equation}
0 \leq P_{eq} \leq \frac{P_t}{N}
\end{equation}
To satisfy the stochastic rate constraint, we plugin $P_i=P_{eq}$ in \eqref{primthput} and solve it for $\bar{R}_{j}^{s} \geq R_j$. Assuming that the solution is $P_{eq} \leq P_{eq}^j \, \forall j \in \{1,2,\dots,M\}$, we can write
\begin{eqnarray}
P_{eq} = \min \left\{\frac{P_t}{N}, P_{eq}^1, P_{eq}^1, \dots, P_{eq}^M\right\}
\end{eqnarray}
Let $R_{k,i}=\log_2 (1+ s_{k,i}P_{eq})$ denote the throughput for the $k^{th}$ SU corresponding to  the $i^{th}$ subcarrier, $A$ denote the set of all subcarriers and $U_k \equiv U_k^{*}$ denote the solution set of subcarriers allocated to $k^{th}$ SU. The algorithm for subcarrier allocation is given in Algorithm 1.
\begin{algorithm}
\begin{enumerate}
\item Initialization \\
set $R_{k,i}\,=\,0$, $A\,=\,\{1,2,...,N\}$ and $U_k^{*}\,=\,\emptyset$ $\forall \, k$
\item For $i \in A$
\begin{description}
\item[(a)] 
find $k \in \{1,2,\dots,K\}$ satisfying $R_{k,i}\,\geq \,R_{k',i}$ $\forall k' \in \{1,2,\dots,K\}-\{k\}$
\item[(b)]
assign $U_k^{*}\,=\, U_k^{*} \cup\, \{i\}$ and $A\,=\,A-\{i\}$.
\end{description}
\item While $A\,\not=\,\emptyset$, repeat step 2.
\end{enumerate}
\caption{Subcarrier Allocation}
\end{algorithm}
Once the subcarriers have been allocated, the optimization problem mentioned in \eqref{jointoptprob} can be restated as,
\begin{equation*}
\max_{P_i} \sum_{k=1}^K \sum_{i \in U_k^{*}} \log_2 (1+s_{k,i} P_i)
\end{equation*}
subject to
\begin{equation*}
\bar{R}_{j}^{s} \geq R_j, \forall j \in \{1,2,\dots,M\} 
\end{equation*}
\begin{equation*}
\sum_{i=1}^N P_i \leq P_t 
\end{equation*}
\begin{equation} \label{P1}
P_i\geq 0, \forall i \in \{1,2,\dots,N\}
\end{equation}

The above mentioned problem is a non-convex optimization problem since the stochastic rate constraint is non-convex. Hence, if we try solving this problem by forming its Lagrange dual problem, then the duality gap between the solutions of primal and its dual problem is non-zero. However, if the optimization problem in \eqref{P1} satisfies the "time-sharing" condition given in \cite{DUAL06}, the duality gap is shown to be zero. In the following lemma, we prove that \eqref{P1} does satisfy the "time-sharing" condition, though with a prerequisite.
\begin{lemma}
The optimization problem mentioned in \eqref{P1} satisfies the "time-sharing" condition when $|\Omega_j| \rightarrow \infty$.
\end{lemma}
\begin{proof}
Please refer to the Appendix.
\end{proof}
The Lagrangian for the optimization problem in \eqref{P1} can be written as,
\begin{multline} \label{lagrangian}
L(\mathbf{P},\lambda,\boldsymbol{\mu})=\sum_{k=1}^K \sum_{i \in U_k^{*}} \log_2 \left(1+s_{k,i}P_i)\right) - \lambda(\sum_{i=1}^N P_i-P_t) \\ -\sum_{j=1}^M \mu_j (R_j-R_j^s)
\end{multline}
where $\lambda$ and $\boldsymbol{\mu}\,=\,[\mu_1,\mu_2,\dots,\mu_M]$ are Lagrangian multipliers and $\mathbf{P}\,=\,[P_1,P_2,\dots,P_N]$ is the power allocation vector.
The corresponding Lagrange dual function is,
\begin{equation}
d(\lambda,\mu)=\max_{\mathbf{P}} L(\mathbf{P},\lambda,\boldsymbol{\mu})
\end{equation}
Thus, the dual optimization problem can be expressed as,
\begin{equation}
\min d(\lambda,\boldsymbol{\mu})
\end{equation}
s.t.
\begin{equation}
\lambda \leq 0,\, \boldsymbol{\mu} \preceq 0.
\end{equation}
The above problem can be solved using dual decomposition method \cite{BOYD}.
\begin{theorem}
The optimal power allocation for the optimization problem in \eqref{P1} is given by,
\begin{equation*}
P_i^*=max \{\beta_o ,0\}
\end{equation*}
where $i \in U_k$ and $\beta_o$ is a positive root of the equation,
\begin{equation*}
\beta = \frac{1}{\lambda +\mu_j v_i(\beta)}-\frac{\Gamma(N_o+J_{k,i})}{|H_{k,i}^{ss}|^2 (1-\alpha_k)}
\end{equation*}
\end{theorem}
\begin{proof}
Please refer to the Appendix.
\end{proof}

If $\lambda$ and $\mu$ are fixed, $\beta_o$ can be found using bisection search \cite{BOYD}. The steps of algorithm for finding optimal sub-carrier power allocation are given in Algorithm 2.
\begin{algorithm}
\begin{enumerate}
\item Initialize $\lambda_1$, $k=1$.
\item Repeat the following steps
\begin{description}
\item[(a)] Initialize $\mu_{j,1}$, $k^{'}=1 \,\forall \, j \,\in \, \{1,2,..M\} \,$
\item[(b)] Repeat the following steps for all $j \in \{1,2,..M\}$
\begin{description}
\item[(i)]
Find $P_i^* \,\forall \, i \,\in \, \Omega_j$ by bisection search
\item[(ii)]
Update $\mu_{j,{k^{'}}}$ by
\begin{equation*}
\mu_{j,{k^{'}+1}}=\mu_{j,{k^{'}}}+\gamma\left(R_j-R_j^s\right)
\end{equation*}
\item[(iii)]
If $\mu_{j,{k^{'}+1}}\leq 0$, put $\mu_{j,{k^{'}+1}}=0$ and stop;
Otherwise stop when $|\mu_{j,{k^{'}+1}}-\mu_{j,{k^{'}}}|\leq \epsilon$.
\end{description}
\item[(c)]
Update $\lambda_{k+1}$ by
\begin{equation*}
\lambda_{k+1}=\lambda_{k}+\eta(\sum_{i=1}^N P_i-P_t)
\end{equation*}
\end{description}
\item If $\lambda_{k+1}\leq 0$, put $\lambda_{k+1}=0$ and stop;
Otherwise stop when $|\lambda_{k+1}-\lambda_{k}|\leq \epsilon$.
\end{enumerate}
\caption{Optimal Power Allocation}
Here $\gamma$ and $\eta$ are step sizes and $\epsilon>0$ is a small constant.
\end{algorithm}


Now, the optimal number of bits allocated to the $i^{th}$ subcarrier, $b_{k,i}^*\, \forall i \in U_k, k \in \{1,2,\dots,K\}$ can be calculated from the optimal power allocation $P_i^*$ using \eqref{bitloadingexp}.
Since the bits allocated to a sub-carrier can only be an integer, $b_i^*$ is rounded off to the next highest integer. But this may increase the total power above $P_t$ which defeats the whole purpose. Hence, we adopt a greedy bit removal algorithm till system constraints are satisfied. The power saved by removing one bit from the $i^{th}$ subcarrier can be obtained from \eqref{bitloadingexp} as, 
\begin{equation}
\Delta P_i=\frac{2^{b_{k,i}-1}}{s_{k,i}}
\end{equation}
We run a greedy bit removal algorithm, shown in Algorithm 3, that removes at each step the bit which recovers maximum power from the subcarriers till the total power reaches below $P_t$.

\begin{algorithm}

\While{$\sum_{i=1}^N P_i>P_t$}{
find $j\in\{1,...,N\}$ such that $\Delta p_j = \max_{i:b_{k,i}>0} \Delta P_i$\;
set $b_j=b_j-1$\;
}
\caption{Greedy bit removal algorithm}
\end{algorithm}


%

\section{Appendix}
\subsection{Proof of Lemma 1}
Let $P_x$ and $P_y$ be the optimal power allocations for the optimization problem in (13) with  maximum rates achieved as $C(P_x)$ and $C(P_y)$, and rate loss constraints as $R_{j,x}^s$ and $R_{j,y}$, respectively. Then in order to show that (13) satisfies the time-sharing condition, we need to prove that for any $0 \leq \theta \leq 1$, there exists a solution $P_z$ for (13), such that $R_{j,z}^s \geq \theta R_{j,x} + (1-\theta)R_{j,y}$ and $C(P_z) \geq \theta C(P_x) + (1-\theta)C(P_y)$. Now, if we construct $P_z$ such that $P_z = P_x$ in $\theta$ fraction of the total set of subcarriers and $P_z = P_y$ in the remaining $(1-\theta)$ fraction, then it is can be easily inferred that $C(P_z) = \theta C(P_x) + (1-\theta)C(P_y)$, so the first condition is satisfied. Moving to the second condition, we can write
\begin{equation}
R_{j,x}^{s}=\sum_{k \in \Psi} \sum_{i \in \Omega_j} \log_2 \left(1+\frac{(|H_i^{pp}| \sqrt{T_i}+|H_{k,i}^{sp}| \sqrt{\alpha_k P_{x,i}})^2}{N_o+|H_{k,i}^{sp}|^2 (1-\alpha_k) P_{x,i}}\right)
\end{equation}

\begin{equation}
R_{j,y}^{s}=\sum_{k \in \Psi} \sum_{i \in \Omega_j} \log_2 \left(1+\frac{(|H_i^{pp}| \sqrt{T_i}+|H_{k,i}^{sp}| \sqrt{\alpha_k P_{y,i}})^2}{N_o+|H_{k,i}^{sp}|^2 (1-\alpha_k) P_{y,i}}\right).
\end{equation}
Based on construction of the power allocation $P_z$, $R_{j,z}^{s}$ can be expressed as
\begin{multline}
R_{j,z}^{s}=\sum_{k \in \Psi} \sum_{i \in \theta \Omega_j} \log_2 \left(1+\frac{(|H_i^{pp}| \sqrt{T_i}+|H_{k,i}^{sp}| \sqrt{\alpha_k P_{x,i}})^2}{N_o+|H_{k,i}^{sp}|^2 (1-\alpha_k) P_{x,i}}\right)+ \\ \sum_{k \in \Psi} \sum_{i \in (1-\theta)\Omega_j} \log_2 \left(1+\frac{(|H_i^{pp}| \sqrt{T_i}+|H_{k,i}^{sp}| \sqrt{\alpha_k P_{y,i}})^2}{N_o+|H_{k,i}^{sp}|^2 (1-\alpha_k) P_{y,i}}\right)
\end{multline}
By law of large numbers, it can be inferred that as $|\Omega_j| \rightarrow \infty$, $R_{j,z}^s = \theta R_{j,x}^s + (1-\theta)R_{j,y}^s$. Since $R_{j,x}^s \geq R_{j,x}$ and $R_{j,y}^s \geq R_{j,y}$, it is easy to prove that $R_{j,z}^s \geq \theta R_{j,x} + (1-\theta)R_{j,y}$. Hence, the second condition for time sharing also holds.
 
\subsection{Proof Theorem 1}
Rewriting \eqref{lagrangian}, we get
\begin{multline}
L(\mathbf{P},\lambda,\boldsymbol{\mu})=\sum_{j=1}^M \sum_{\substack{i \in \Omega_j \cap U_k \\ k \in \{1,2,\dots,K\}}}\log_2 (1+s_{k,i}P_i) \\ - \lambda( \sum_{j=1}^M \sum_{i \in \Omega_j} P_i) - \sum_{j=1}^M \mu_j (R_j-R_j^s)+\lambda P_t
\end{multline}
Accordingly, the Lagrangian dual function can be expressed as,
\begin{equation} \label{dualfunc1}
d(\lambda,\boldsymbol{\mu})=\sum_{i=1}^M d_j^{'}(\lambda,\boldsymbol{\mu})+\lambda P_t
\end{equation}
where,
\begin{multline}
d_j^{'}(\lambda,\boldsymbol{\mu})=\max_{P_i \in \Upsilon_j} \sum_{\substack{i \in \Omega_j \cap U_k \\ k \in \{1,2,\dots,K\}}}\log_2 (1+s_{k,i}P_i) \\ - \lambda \sum_{i \in \Omega_j} P_i-\mu_j (R_j-R_j^s)
\end{multline}
and $\Upsilon_j$ is defined as $\Upsilon_j = \{P_i : P_i \geq 0, \forall i \in \Omega_j\}$. It can be inferred that for a given $\lambda$, \eqref{dualfunc1} can be decomposed into $M$ independent optimization problems, each having the description,

\begin{equation*}
\max_{\mathbf{P_j}} z(\mathbf{P_j})
\end{equation*}
subject to
\begin{equation} \label{P2}
R_j^s \geq R_j
\end{equation}
where 
\begin{equation}
z(\mathbf{P_j})=\sum_{\substack{i \in \Omega_j \cap U_k \\ k \in \{1,2,\dots,K\}}}\log_2(1+s_{k,i}P_i) - \lambda \sum_{i \in \Omega_j} P_i
\end{equation}

The Lagrangian for the optimization problem in \eqref{P2} can be written as,
\begin{equation}
L_j^{'}(\mathbf{P_j},\mu_j)=z(\mathbf{P_j})-\mu_j(R_j-R_j^s)
\end{equation}
The Lagrange dual function can then be expressed as,
\begin{equation} 
d^{''}(\mu_j)=\max_{\mathbf{P_j}} L_j^{'}(\mathbf{P_j},\mu_j)
\end{equation}
Thus the dual optimization problem becomes,
\begin{equation*}
\min_{\mu_j} d^{''}(\mu_j)
\end{equation*}
s.t.
\begin{equation} \label{P3}
\mu_j \leq 0.
\end{equation}

Hence the KKT conditions for \eqref{P3} are,
\begin{equation}
\mu_j(R_j-R_j^s)=0
\end{equation}
\begin{equation}
\frac{\partial L_j^{'}(\mathbf{P_j},\mu_j)}{\partial P_i}=0
\end{equation}

From the KKT conditions given above, the optimal power allocation satisying \eqref{P2} can be found as follows.
\begin{multline} \label{lagrangediff}
\frac{\partial L_j^{'}(\mathbf{P_j},\mu_j)}{\partial P_i}=\frac{|H_{k,i}^{ss}|^2(1-\alpha_k)}{\Gamma(N_o+J_i)+|H_{k,i}^{ss}|^2 P_i(1-\alpha_k)} \\ +\mu_j \frac{\partial R_j^s}{\partial P_i}
\end{multline}
where $i \in \Omega_j \cap U_k$ and $k \in \{1,2,\dots,K\}$.
Now,
\begin{equation} \label{viPi}
\frac{\partial R_j^s}{\partial P_i}= \frac{N}{D_1D_2}
\end{equation}
where,
\begin{multline}
N = N_o (|H_{k,i}^{sp}|^2 \alpha_k + |H_i^{pp}| \sqrt{T_i}|H_{k,i}^{sp}| \sqrt{\alpha_k /P_i}) \\  - |H_i^{pp}|^2 |H_{k,i}^{sp}|^2 T_i(1-\alpha_k) \\ + |H_{k,i}^{sp}|^2(1-\alpha_k)P_i|H_i^{pp}| \sqrt{T_i}|H_{k,i}^{sp}| \sqrt{\alpha_k /P_i} \\  - 2|H_{k,i}^{sp}|^2(1-\alpha_k)|H_i^{pp}| \sqrt{T_i}|H_{k,i}^{sp}| \sqrt{\alpha_k P_i},
\end{multline}
\begin{equation}
D_1 = (N_o+|H_{k,i}^{sp}|^2(1-\alpha_k)P_i),
\end{equation}
\begin{multline}
D_2 = (N_o+|H_{k,i}^{sp}|^2(1-\alpha_k)P_i \\ + (|H_i^{pp}| \sqrt{T_i}+|H_{k,i}^{sp}| \sqrt{\alpha_k P_i})^2).
\end{multline}
Expressing \eqref{viPi} as $-v_i(P_i)$ and equating \eqref{lagrangediff} to zero, we get
\begin{equation}
P_i=\frac{1}{\lambda +\mu_j v_i(P_i)}-\frac{\Gamma(N_o+J_{k,i})}{|H_{k,i}^{ss}|^2 (1-\alpha_k)}
\end{equation}

\bibliography{biblio}
\bibliographystyle{ieeetr}

\end{document}